\documentclass[a4paper,11pt]{article}
\usepackage{fullpage}
\usepackage{times}
\usepackage{subfigure}
\usepackage{vmargin}
\usepackage{fancyheadings}
\usepackage[cp1251]{inputenc}
\usepackage[english]{babel}
\usepackage{graphicx}
\usepackage{color}
\usepackage{cite}
\usepackage{amsmath,amssymb,amsthm,enumerate,bbm,color,verbatim,setspace}
\newcommand{\tr}{\text{Tr}}
\usepackage{braket}
\theoremstyle{plain}  
\newtheorem{thm}{Theorem}[section]

\newtheorem{cor}{Corollary}[section]

\newtheorem{exm}{Example}[section]

\theoremstyle{remark}

\date{}
\title{Not all Probability Density Functions are Tomograms}
\author{L.A. Markovich$^{1,2,3,\dagger}$, J. Urbanetz$^{1}$  and V.I. Man'ko$^{3,4}$\\
     $^{1}$ \quad  Institute-Lorentz, University Leiden, P.O. Box 9506, \\2300 RA Leiden, The Netherlands \\
$^{2}$ \quad Institute for information transmission problems,  Bol. Karetny per. 19,\\ Moscow 127051, Russia\\
$^{3}$ \quad Russian Quantum Center, Skolkovo, Moscow 121205, Russia\\
$^{4}$ \quad Lebedev Physical Institute, Russian Academy of Sciences, \\Leninskii Prospect 53, Moscow 119991, Russia\\
$^*$Corresponding author e-mail: markovich@mail.lorentz.leidenuniv.nl}
\date{}
\begin{document}
\maketitle
\pagenumbering{arabic}
\begin{abstract}\noindent

This paper delves into the significance of the tomographic probability density function (pdf) representation of quantum states, shedding light on the special classes of pdfs that can be tomograms. 
Instead of using wave functions or density operators on Hilbert spaces, tomograms, which are the true pdfs, are used to completely describe the states of quantum systems.  Unlike quasi-pdfs, like the Wigner function, tomograms can be analysed using all the tools of classical probability theory for pdf estimation, which can allow a better quality of state reconstruction. This is particularly useful when dealing with non-Gaussian states where the pdfs are multi-mode. The knowledge of the family of distributions plays an important role in the application of both parametric and non-parametric density estimation methods.  We show that not all pdfs can play the role of tomograms of quantum states and introduce the conditions that must be fulfilled by pdfs to be "quantum". %The latter conditions are fulfilled by the tomograms of the continues states like oscillators, cats states and their superposition. 
\end{abstract}

\medskip

\section{Introduction}
 \par
In the early 1900s, Schr{\"o}dinger introduced the concept of a quantum system's state~\cite{schrodinger1977quantisierung}, characterised by a complex wave function. 
 The notion of a density matrix~\cite{landau1927dampfungsproblem,neumann1927wahrscheinlichkeitstheoretischer,dirac1981principles} or a density operator, acting in a Hilbert space, was established shortly after to characterise the generic states of quantum systems.
\par Nowadays, quantum state reconstruction is a critical task in quantum information science and quantum technology. 
%It entails the estimation of the quantum state from measurements in the presence of noise that can affect the accuracy and reliability of the reconstructed state. 
We know that in classical physics, it is always possible to assign a unique probability distribution function (pdf) to the outcomes of a physical measurement in continuous-variable systems. Based on the measurement results, one can estimate the pdf using different parametric and nonparametric methods~\cite{silverman1986density} and thus fully describe the system. However, in quantum mechanics, the Heisenberg uncertainty principle states that certain pairs of non-commuting observables, such as position and momentum, cannot be simultaneously precisely determined. 
Therefore, it is challenging to reconstruct the quantum state from measurements, which leads to the need for a more sophisticated representation of the quantum state.
\par Quasi-probability distributions (qpdf) are introduced to represent the probabilities of outcomes of non-commuting observables, such as position and momentum, in a way that respects the uncertainty principle.
Qpdfs are not actual probabilities in the classical sense. These distributions are often complex-valued and can take negative values, which can make their interpretation challenging. 
%However, they provide a valuable tool for describing and analyzing quantum states and phenomena.
The Wigner function is one of the most well-known qpdfs~\cite{PhysRev.40.749}. It provides a phase-space representation of a quantum state, allowing the simultaneous description of position and momentum. The Wigner function is real-valued and can take negative values.
The Husimi Q-function is another qpdf~\cite{1940264,doi:10.1063/1.1704739} that is positive and can be thought of as a smoothed version of the Wigner function~\cite{linowski2023relating}. The Glauber-Sudarshan P-function~\cite{PhysRevLett.10.84,PhysRevLett.10.277} is used to represent states in the context of quantum optics and can be used for the description of coherent states.
There exist other quasi-probabilities defined as different Fourier-like transforms, like Kirkwood-Rihaczek (KR) \cite{PhysRev.44.31}, Generalized KR~\cite{PRAXMEYER2003349}, Margenau-Hill (MH)~\cite{10.1143/PTP.26.722}, Page~\cite{doi:10.1063/1.1701949}, Choi-Williams (CW)~\cite{28057}, sinc~\cite{30749}, and Cohen~\cite{doi:10.1063/1.1931206}.

A symplectic tomogram~\cite{MANCINI19961,Mancini_1995} is an alternative way to represent the quantum state. %, especially when dealing with non-Gaussian states. 
 It is a real-valued continuous probability distribution in the phase space and is constructed through a convolution of the density operator with a kernel that depends on the phase space coordinates. The resulting function represents the probability of finding the quantum system in a particular position and momentum configuration. The symplectic tomogram is  related to the qpdfs with the invertible integral transforms~\cite{man2020integral}.  In fact, the Wigner function can be derived from the symplectic tomogram through an inverse Radon transform~\cite{Ibort_2009}. 

Since the symplectic tomogram is a pdf, one can use all the techniques known in probability theory for pdf estimation. 
The parametric pdf estimation methods strongly rely on the knowledge of the class of the pdf. For example, the histogram method is widely used in practise~\cite{vrehavcek2008tomography,fabre2023local} because of its simplicity. It divides the data range into intervals or bins and counts the number of data points in each bin. The pdf is then approximated by normalizing these counts. However, in some cases, a data sample does not resemble a common probability distribution or cannot be easily made to fit the distribution. 
This is often the case when the data has two peaks (bimodal distribution) or many peaks (multi-modal distribution). In this case, the  nonparametric methods are used. These methods do not assume a specific parametric form for the pdf. 
%Perhaps the most common nonparametric approach for estimating the probability density function of a continuous random variable is called kernel density estimation (KDE). In addition to KDE, nonparametric methods include Parzen Windows, Gaussian Mixture Models, and Locally Weighted Scatterplot Smoothing.
%As we can see, the knowledge of the distribution class of the tomogram is essential for its reconstruction. 
Hence, knowledge of the pdf family is essential for an optimal selection of the pdf reconstruction method.
\par By definition, a state is called Gaussian if its Wigner function is a Gaussian like function~\cite{walschaers2021non}. 
Since the Wigner function and the tomogram are connected by the Radon transform, we can say that the state is Gaussian if its tomogram is a Gaussian pdf. 
In~\cite{man1997quantum} the  quantum  harmonic oscillator (HO) is studied, showing that its tomograms are  from the  exponential family of distributions~\cite{barndorff2014information}. In particular, the tomogram of the ground state  is the Gaussian pdf.  The inverted oscillator~\cite{man2023inverted} state tomogram is also from the exponential family. 
The tomograms are  known to be Gaussian pdfs  for the squeezed vacuum states and the thermal
states~\cite{man2023quantum}. 
However, a superposition of Gaussian states
 is already a non-Gaussian mixture~\cite{MoyaCessa2008OpticalPO}. There has been a
strong experimental focus on two specific types of non-Gaussian states:
 cat states~\cite{yurke1986generating} and Gottesman-Kitaev-Preskill (GKP)~\cite{gottesman2001encoding} states.
  In~\cite{lopez2022symplectic} the crystallized cat states tomograms are presented. Being the superposition of  cat states, their states result in being  non-Gaussian too since their tomogram is the sum of Gaussian  functions, which in general is not a Gaussian pdf.

Thus, non-Gaussian states are frequently encountered in practice. However, can entirely different distributions describe a quantum state, for instance, from the exponential or power-law family?
It is evident that not all pdfs are suitable for the role of a tomogram due to the constraints imposed on the quantum state density operator, e.g., that it is nonnegative, Hermitian, and has a unit trace. Thus, the question we address in this paper is: What continuous distributions can describe quantum states? Knowing the answer, we can solve the inverse problem: given a pdf playing the role of the tomogram, find the new corresponding quantum state not known in the literature.

 \subsection{Contributions of this paper}
\par In this manuscript, we introduce the characteristic function of the tomogram pdf as a description of the quantum state. Being the Fourier transform of the pdf, the characteristic function always exists and is bounded. We 
%\textcolor{blue}{when writing have shown here, don't write we show later. I think I would stick with "we show"}
show that the characteristic function of the pdf corresponding to the quantum state must fulfil the conditions following from the density operator properties. If the distribution does not satisfy these conditions, then the corresponding quantum state does not exist. 
%\textcolor{blue}{Thus, it is possible to assess the ability of known/in literature known pdfs to generate physical quantum states} 
Thus, it is possible to assess the ability of known pdfs to generate physical quantum states.
\par The trace of the density operator and the purity parameter are explicitly written in terms of the characteristic function. Moreover, we show that the trace of the product of two density operators can be written as an integral from their characteristic functions corresponding to their tomogram pdfs. Similarly the fidelity parameter is written. 
Thus, knowledge about the characteristic function can be used for purity testing in the case of two identical states or in the case of two different states for testing their overlap.
Also, the elements of the density matrix are written in terms of the characteristic function, giving a novel way of parametrizating the density matrix.
\par We investigate the exponential family of pdfs to find the conditions on its parametrization to satisfy the constraints on the characteristic function. We check some known pdfs from the exponential family: exponential, gamma, and $\chi^2$ pdfs, showing that they never satisfy the latter conditions, thus not being suitable for quantum state generation. 
\par Finally, we study particular systems with continuous variables like oscillators, the superposition of cat states, and different Gaussian states. Their pdfs either belong to the exponential family of pdfs or are a mixture of Gaussians. The corresponding characteristic functions are derived, showing that they satisfy the conditions imposed on the characteristic function to generate a physical quantum state. For a pseudoharmonic oscillator~\cite{olendski2023one} that is defined on the positive semiaxis, we deduce the tomogram for the first time, showing it is from the exponential family too. 
\subsection{Organization of the paper}
\par The paper is organised as follows. In Sec.~\ref{sec_1} we recall the notion of the symplectic tomogram, giving its connection to the density matrix of the quantum state. In Sec.~\ref{sec_2} the characteristic function of the tomogram is introduced. The requirements that must be satisfied to characterize a quantum state are outlined. In Sec.~\ref{sec_4} the general exponential family of pdfs is considered. We study the examples of continuous pdfs known in classical probability theory. The quantum harmonic and pseudoharmonic oscillators  are studied in detail in Sec.~\ref{sec_3}.
The superposition of cat states is analysed in Sec.~\ref{sec_5}.  

\section{Symplectic Tomogram}\label{sec_1}
\par  
Let us consider a quantum state in an infinite-dimensional Hilbert space $\mathcal{H}$ associated with a positive Hermitian operator $\hat{\rho}$, called the density matrix. The kernel of the density matrix in coordinate representation is $\rho(x,x')=\bra{x}\rho\ket{x'}$, where  $\ket{x}$ is an eigensate of the position operator $\hat{q}$.
The density matrix operator is hermitian ($\rho^{\star}(x,x')=\rho(x',x)$). Its diagonal elements are nonnegative ($\rho(x,x)\geq 0$) and its trace is equal to one ($\int \rho(x,x)dx=1$). \par According to the general scheme~\cite{MANCINI19961}, in the case of continuous variables, the mapping of the density matrix to the family of pdfs, depending on two real parameters $\mu$ and $\nu$, is given by the following relation:
\begin{eqnarray}\label{1}
\mathcal{W}(X|\mu,\nu)=\tr{(\hat{\rho}\delta\left(\hat{1} X-\mu \hat{q}-\nu\hat{p}\right))}=\bra{X;\mu,\nu}\hat{\rho}\ket{X;\mu,\nu}.
\end{eqnarray}
The $\delta(\cdot)$ is the Dirac delta-function.
Here    $\hat{q}$ and $\hat{p}$  are the position and momentum operators:
\begin{eqnarray}
    \hat{q}\ket{X;\mu,\nu}=X\ket{X;\mu,\nu}, 
\end{eqnarray}
 and $|X;\mu,\nu\rangle$ is an eigenvector of the hermitian operator 
\begin{eqnarray}
    \hat{X}(\mu,\nu)=\mu \hat{q}+\nu \hat{p},\quad \hbar=1. 
\end{eqnarray}
The latter is a canonical transform of $\hat{q}$ and $\hat{p}$. Formally, this quantity
 is a coordinate, measured in scaled and rotated reference frame in the phase space. 
Its  pdf \eqref{1} is called the \textit{symplectic tomogram}. 
As a pdf the tomogram is nonnegative and normalised:
\begin{eqnarray}
\int_{-\infty}^{\infty}\mathcal{W}(X|\mu,\nu)dX=1.
\end{eqnarray}
 For the symplectic tomogram, the inverse
quantum Radon transform is the following:
\begin{eqnarray}\label{1311}
    \hat{\rho}=\frac{1}{2\pi}\int \mathcal{W}(X|\mu,\nu) \exp{(i(X\hat{1}-\mu\hat{q}-\nu\hat{p}))}dXd\mu d\nu,
\end{eqnarray}
defining the density matrix operator by the corresponding pdf. 
For the density operator $\hat{\rho}=\ket{\psi}\bra{\psi}$ of the pure state $\ket{\psi}$ the relation \eqref{1} converts into
\begin{eqnarray}\label{1645_5}
    \mathcal{W}(X|\mu,\nu)=\frac{1}{2\pi|\nu|}\Big|\int\limits_{-\infty}^{\infty} \Psi(y)\exp{(\frac{i\mu}{2\nu}y^2-\frac{iX}{\nu}y)} dy\Big|^2.
\end{eqnarray}
%where $\Psi(y)$ is the wave function of the state.
\par The representation \eqref{1311} is connected with the Weyl and star-product quantization~\cite{ibort2009introduction}. It is known that the Weyl symbol of the density matrix is explicitly the Wigner
function. It is shown in~\cite{mancini1995wigner} that the symplectic tomogram is related to the quantum state expressed in terms of its Wigner function $W(q,p)$ with an integral transform.
%as follows:
%\begin{eqnarray}\label{1720}
%   \mathcal{W}(X|\mu,\nu)=\frac{1}{2\pi}\int \limits_{-\infty}^{\infty} W(q,p)\delta{(X-\mu q-\nu p)} dq dp.
%\end{eqnarray}
The parameters $\mu$ and $\nu$  describe an ensemble of rotated
and scaled reference frames, in which the observable $X$ is measured.  For $\mu=\cos{\phi}$ and $\nu=\sin{\phi}$,
the pdf $\mathcal{W}(X|\mu,\nu)$ is the distribution for homodyne output variable used in optical
tomography~\cite{PhysRevA.40.2847,leonhardt1994high}. The procedure of balanced homodyne photon detection is based on mixing of a measurable (weak)
field and a strong coherent field with varying phase $\phi$ on the beam splitter. In this case, the measurable
observable is $\hat{X}=\hat{q}\cos{\phi}+\hat{p}\sin{\phi}$. The angle $\phi$ could be interpreted as a rotation angle of the phase space. 
 Relations \eqref{1}-\eqref{1645_5} for symplectic tomogram are transformed to equivalent relations for optical tomograms.  Note that the symplectic tomogram
is a function of two parameters $(\mu,\nu)$ of the $Sp(2,\mathrm{R})$ group parameterization, and the optical tomogram is a function of the parameter $\phi$. 
\par Alternatively, the parameters $\mu$ and $\nu$ can be expressed in the form $s \cos{\phi}$, $s^{-1}\sin{\phi}$, where $s > 0$ is a real squeezing parameter and $\phi$ is a rotation angle.  
Then the variable $X$ is identical to the position measured in the new reference frame in the phase-space with axis $sq$ and $s^{-1}p$ and after the scaling the axis are rotated by an angle $\phi$. Thus the tomogram implies the pdf  of the random position $X$ measured in the new (scaled and rotated) reference frame in the phase-space~\cite{del2008symplectic}. 
%\par 
%Formula \eqref{1720} can be inverted and the Wigner function of the state can be
%expressed in terms of the pdf as follows
%\begin{eqnarray}\label{1844}
%W(q,p)  =\frac{1}{2\pi}\int \limits_{-\infty}^{\infty}    \mathcal{W}(X|\mu,\nu) \exp{(-i(\mu q+\nu p-X))} d\mu d\nu dX.
%\end{eqnarray}
%Note that in the classical case, $W(q,p)$ would be a pdf. However, in the quantum case $W(q,p)$ can assume negative values, so  we call it a quasi-pdf.
%Given a set of distribution functions
%$\mathcal{W}(X|\phi)$ for quadrature values $X$ with $\phi$ gradually
%varying from $0$ to $\pi$, the quantum Wigner function of the
%mode can be reconstructed even for mixed states~\cite{vogel1989determination}.
\par The process of inverting the raw observed data in order to arrive at a form of the quantum state is both delicate and intriguing. Deterministic and nondeterministic inversion methods are the two primary categories of inversion techniques. In the deterministic methods  an experimentally determined pdf (tomogram) is used to
determine the matrix elements of the density matrix, by a direct mathematical inversion of \eqref{1311}. However this strongly depends on how well we estimated the pdf based on the measurement of $X$ for varying $\phi$.
The nondeterministic technique aims to directly estimate the quantum state, as opposed to using classical distributions as an  intermediate object~\cite{jones1994fundamental,schack2001quantum}.
\section{Characteristic Function}\label{sec_2}
\label{sec charac}
Since the tomogram is the pdf of the variable $X$ depending on real parameters $\mu$ and $\nu$, we can introduce its  characteristic function. According to the definition the characteristic function
\begin{eqnarray}
    \phi_X(t)\equiv \int\limits_{-\infty}^{\infty}\mathcal{W}(X)e^{itX}dX
    \label{eqn pdf to charac}
\end{eqnarray}
 is the Fourier transform of the pdf. It is known that a random variable $X_n$ weakly converges to a random variable $X$ if and only if, for any $t$ the characteristic function $\phi_{X_n}(t)$ converges to the characteristic function $\phi_{X}(t)$~\cite{kolmogorov2018foundations}. Further, we omit $X$ in the notation $\phi_X(t)$.
 \par  
The characteristic function of any real-valued random variable completely defines its probability distribution. According to definition, it is non-vanishing in a region around zero ($\phi(0) = 1$) and it is bounded $|\phi(t)| \leq 1$, $\forall t$.
We can introduce the characteristic function $\phi(1;\mu,\nu)$ at point $t=1$ corresponding to the tomogram pdf \eqref{1}:
 \begin{eqnarray}
    \phi(1;\mu,\nu)\equiv \int\limits_{-\infty}^{\infty}\mathcal{W}(X|\mu,\nu)e^{iX}dX.
    \label{eqn pdf to charac_1}
\end{eqnarray}
Then we can rewrite the density operator \eqref{1311} as follows:
\begin{eqnarray}
    \hat{\rho}=\frac{1}{2\pi}\int  \phi(1;\mu,\nu)e^{-i(\mu\hat{q}+\nu\hat{p})}d\mu d\nu,
\end{eqnarray}
and, taking the  trace from both sides, we can conclude
\begin{eqnarray}
   \tr{ \hat{\rho}}%&=&
   %\frac{1}{2\pi}\int \mathcal{W}(X|\mu,\nu) e^{i X}\tr{(e^{-i(\mu\hat{q}+\nu\hat{p})})}dXd\mu d\nu\\\nonumber
  =
   \int e^{i\frac{\mu\nu}{2}}\phi(1;\mu,\nu)\delta(\mu)\delta(\nu)d\mu d\nu= \phi(1;0,0),
\end{eqnarray}
where we used $Tr(e^{-i(\nu\hat{p}+\mu\hat{q})})=2\pi e^{i\frac{\mu\nu}{2}}\delta(\mu)\delta(\nu)$. Thereby,  the trace of the density matrix of a quantum state is the characteristic function with $\mu=0,\nu=0$. That gives us the condition on the characteristic function of the quantum state: $
    \phi(1;0,0)=1$.
\par Using \eqref{1311}, we can write the product of two density matrices, $\hat{\rho}_1$ and $\hat{\rho}_2$ corresponding to two quantum states.
%can be written as
%\begin{eqnarray}\label{1035}
%   \hat{ \rho}_1\hat{\rho}_2&=& \frac{1}{(2\pi)^2}\iint \mathcal{W}_1(X_1|\mu_1,\nu_1)  \mathcal{W}_2(X_2|\mu_2,\nu_2)\\\nonumber &\times&e^{iX_1}e^{iX_2}dX_1 dX_2 e^{-i(\mu_1 \hat{q}+\nu_1 \hat{p})}e^{-i(\mu_2 \hat{q}+\nu_2 \hat{p})} d\mu_1d\mu_2d\nu_1d\nu_2.
%\end{eqnarray}
Taking the trace, we get
\begin{eqnarray}\label{1036}
  \tr{( \hat{ \rho}_1\hat{\rho}_2)}
  %\nonumber&=& \frac{1}{(2\pi)^2}\iint \phi_1(1;\mu_1,\nu_1) \phi_2(1;\mu_2,\nu_2) \tr{( e^{-i((\mu_1+\mu_2) \hat{q}+(\nu_1+\nu_2) \hat{p})})} d\mu_1d\mu_2d\nu_1d\nu_2\\&=&\frac{1}{(2\pi)} \iint \phi_1(1;\mu_1,\nu_1) \phi_2(1;\mu_2,\nu_2)\delta(\mu_1+\mu_2)\delta(\nu_1+\nu_2)  d\mu_1d\mu_2d\nu_1d\nu_2\\\nonumber
  &=& \frac{1}{2\pi}\iint \phi_1(1;\mu_1,\nu_1) \phi_2(1;-\mu_1,-\nu_1) d\mu_1d\nu_1,
\end{eqnarray}
where $\phi_1(1;\mu_1,\nu_1)$ and $\phi_2(1;\mu_1,\nu_1)$ are the characteristic functions, corresponding to the tomograms of the states. 
\par 
Since the modulus of the characteristic function is bounded by one, we can be sure that the latter integral is also always bounded by one that fully coincides with the left-hand side trace upper bound. However, the integral can be negative since the characteristic function in general can take negative values. The left-hand side trace from the product of two density matrices is always non-negative, so the condition on the characteristic function of the quantum state holds:
\begin{eqnarray}\label{1443}
 0\leq\frac{1}{ 2\pi} \iint \limits_{-\infty}^{\infty} \phi_1(1;\mu,\nu) \phi_2(1;-\mu,-\nu)d\mu d\nu\leq 1.
\end{eqnarray}
Formula \eqref{1036} is an analogy of the density matrix distinguishability test.
\begin{thm}
    Two quantum states with the density matrices $\hat{\rho}_1$ and $\hat{\rho}_2$ are distinguishable if the integral from their characteristic functions \eqref{1036} is different from one. 
\end{thm}
When $\hat{\rho}_1=\hat{\rho}_2$, holds, the left-hand side of \eqref{1036} is the purity parameter 
\begin{eqnarray}\label{1404}
 \tr{(\hat{\rho}^2)} &=&\frac{1}{2\pi} \iint \limits_{-\infty}^{\infty} \phi(1;\mu,\nu) \phi(1;-\mu,-\nu)d\mu d\nu.
\end{eqnarray}
Hence, if one knows the characteristic functions of the states, one can calculate the purity or trace product of the density matrices. 
\par The fidelity of two quantum states can be written in the same way as:
\begin{eqnarray}
    \omega_{\psi_1\rightarrow\psi_2}=\tr{(\rho_{\phi_2}\rho_{\psi_1}^{\dagger})}=
    \frac{1}{2\pi}\int  \phi_2(1;\mu,\nu)\phi_1(-1;-\mu,-\nu)d\mu d\nu.
\end{eqnarray}
\par Since the density matrix is hermitian, its matrix elements satisfy $\rho(y,y')=\rho^{\star}(y',y)$. 
The matrix element of \eqref{1311} is (see the deduction in Appendix~\ref{app_1})
\begin{eqnarray}
    \rho(y,y')=\frac{1}{2\pi}\int \phi(1;\mu,y-y')\exp{\left(-i\frac{\mu(y+y')}{2}\right)}  d\mu.
    \label{eqn: integral from matrix elements}
\end{eqnarray}
We can conclude that for a quantum state, the characteristic function must satisfy:
\begin{eqnarray}
    \phi(1;\mu,y-y')= \phi(-1;\mu,y'-y),\quad \forall y,y'.
\end{eqnarray}
The diagonal elements of the density matrix can be written as
\begin{eqnarray}
    \rho(y,y)=\frac{1}{2\pi}\int \phi(1;\mu,0)\exp{\left(-i\mu y\right)}d\mu\equiv \hat{\phi}(1;y,0),
\end{eqnarray}
where $\hat{\phi}(1;y,0)$ is a Fourier transform of $\phi(1;\mu,0)$. We can conclude that $\hat{\phi}(1;y,0)\geq 0$.
\par 
We can summarise the latter results in the following theorem:
\begin{thm}
\label{thm density matrix}
    The characteristic function of a quantum state defines a density operator $\hat{\rho}$  if it satisfies the following conditions:
 \begin{eqnarray}\label{th1}
\tr{\hat{\rho}}=1:\quad  \phi(1;0,0)=1,
 \end{eqnarray}
  \begin{eqnarray}\label{th2}
0\leq\tr{\hat{\rho}^2}\leq1:\quad 0\leq\quad \frac{1}{2\pi} \iint \limits_{-\infty}^{\infty} \phi(1;\mu,\nu) \phi(1;-\mu,-\nu)d\mu d\nu\leq 1,
 \end{eqnarray}
  \begin{eqnarray}\label{th3}
{\rho}^{\star}(y,y')={\rho}(y',y):\quad  \phi(1;\mu,y-y')= \phi(-1;\mu,y'-y),\quad \forall y,y',
\end{eqnarray}
\begin{eqnarray}
    \rho(y,y)\geq 0:\quad \hat{\phi}(1;y,0)\geq 0,\quad \forall y.
    \label{eqn: integral from matrix elements}
\end{eqnarray}

For two density matrices $\hat{\rho}_{1}$, $\hat{\rho}_{2}$ the following condition 
  \begin{eqnarray}\label{th5}
0\leq \tr{(\hat{\rho}_{1}\hat{\rho}_{2})}\leq 1:\quad  0\leq\frac{1}{ 2\pi} \iint \limits_{-\infty}^{\infty} \phi_1(1;\mu,\nu) \phi_2(1;-\mu,-\nu)d\mu d\nu\leq 1,
 \end{eqnarray}
 holds. 
\end{thm}
\par Hence, if we want to reconstruct a state from a given tomographic pdf, we need to ensure that its characteristic function satisfies the Theorem~\ref{thm density matrix}. Next, we  consider several examples.
\subsection{Harmonic oscillator}\label{ex_2}
The ground state of the harmonic oscillator (HO) is defined by the wave function
\begin{eqnarray}
   \Psi_0(y)=\pi^{-1/4}\exp{(-y^2/2)}.
\end{eqnarray}
The corresponding tomogram can be easily calculated
\begin{eqnarray}\label{1004}
     \mathcal{W}_0(X|\mu,\nu) =
     \frac{1}{\sqrt{\pi(\mu^2+\nu^2)}}\exp{\left[-\frac{X^2}{\mu^2+\nu^2}\right]}.
\end{eqnarray}
The latter is a Gaussian pdf  $N(0,\frac{\mu^2+\nu^2}{2})$.
The corresponding characteristic function is
\begin{eqnarray}\label{1407}
    {\phi}_0(1;\mu,\nu)=\exp{\left[-\frac{\mu^2+\nu^2}{4}\right]},
\end{eqnarray}
and it is resistant to the change of the sign of $\mu$, $\nu$ of the parameters.
%Then the matrix element of the density matrix is
%\begin{eqnarray} \Big[\hat{\rho}\Big]_{yy'}&=&
   %\frac{1}{2\pi}\int   \exp{\left[-\frac{\mu^2+\nu^2}{4}\right]}e^{-i\mu x} \delta(y-x)\delta(x-(y'+\nu))\exp{(i\frac{\mu\nu}{2})}d\mu d\nu dx\\\nonumber
  % &=& \frac{1}{2\pi}\int  \exp{\left[-\frac{\mu^2+\nu^2}{4}\right]} e^{-i\mu y}\delta(y-(y'+\nu))\exp{(i\frac{\mu\nu}{2})}d\mu d\nu \\\nonumber
 %  &=&  \frac{1}{2\pi}\int  \exp{\left[-\frac{\mu^2+(y'-y)^2}{4}\right]} e^{-i\mu y}\exp{(-i\frac{\mu(y'-y)}{2})}d\mu  \\\nonumber
 %  &=&\frac{1}{\sqrt{\pi}}\exp{\left[-\frac{y^2+y'^2}{2}\right]}.\end{eqnarray}
%That coincides with the product of two wave functions of the harmonic oscillator at temperate zero.
It is easy to check that the Theorem~\ref{thm density matrix} is fulfilled.
\par 
The wave function of the excited state of the harmonic oscillator is
\begin{eqnarray}\Psi_n(y)=\frac{1}{\sqrt[4]{\pi}\sqrt{2^nn!}}e^{-y^2/2}H_n(y),\end{eqnarray}
where $H_n(y)$ is the Hermite polynomial.
We use the integral 
\begin{eqnarray}\label{1633}
    \int\limits_{-\infty}^{\infty}e^{-p(x-y)^2}H_n(cx)dx=\frac{\sqrt{\pi}(p-c^2)^{n/2}}{p^{(n+1)/2}}H_n\left(cy\sqrt{\frac{p}{p-c^2}}\right)
\end{eqnarray}
to find the tomogram of the excited oscillator state:
\begin{eqnarray}\label{1100}
     \mathcal{W}_n(X|\mu,\nu) =
    \mathcal{W}_0(X|\mu,\nu)\frac{1}{2^n n!}\left(H_n\left(\frac{X}{\sqrt{\mu^2+\nu^2}}\right)\right)^2.
\end{eqnarray}
To our best knowledge, the latter pdf is not arising in classical probability theory in any context except quantum mechanics. 
In some rare sources, it is named the Hermite-Gaussian pdf. 
We use the following integral:
\begin{eqnarray}
    \int\limits_{-\infty}^{\infty}e^{-z^2+bz}H_n(z)H_m(z)dz = \sqrt{\pi}e^{\frac{b^2}{4}}n!m!\sum\limits_{k=0}^{\min{(n,m)}}\frac{2^kb^{n+m-2k}}{k!(n-k)!(m-k)!},
\end{eqnarray}
to find its characteristic function
\begin{eqnarray}
    {\phi}_n(1;\mu,\nu)
    %&\equiv&  \int\limits_{-\infty}^{\infty}\frac{1}{\sqrt{\pi(\mu^2+\nu^2)}}\exp{\left[-\frac{X^2}{\mu^2+\nu^2}+iX\right]}\frac{1}{2^n n!}\Big(H_n\left(\frac{X}{\sqrt{\mu^2+\nu^2}}\right)\Big)^2dX\\\nonumber
   % &=&\frac{1}{2^n n!} \int\limits_{-\infty}^{\infty}e^{-y^2+i\sqrt{\mu^2+\nu^2}y}H_n^2\left(y\right)dy\\\nonumber
   % &=&\frac{n!}{2^n } e^{\frac{-(\mu^2+\nu^2)}{4}}\sum\limits_{k=0}^{n}\frac{2^k(i\sqrt{\mu^2+\nu^2})^{2n-2k}}{k!((n-k)!)^2}\\\nonumber
   % &=&\frac{n!}{2^n } e^{\frac{-(\mu^2+\nu^2)}{4}}\sum\limits_{k=0}^{n}\frac{2^k(-1)^{n-k}(\mu^2+\nu^2)^{n-k}}{k!((n-k)!)^2}\\\nonumber
    &=&e^{\frac{-(\mu^2+\nu^2)}{4}} L_n\left(\frac{\mu^2+\nu^2}{2}\right),
    \label{eqn n-th excited charac}
\end{eqnarray}
where we used the Laguerre polynomial series decomposition
\begin{eqnarray}
    L_n(x)=\sum\limits_{k=0}^n\frac{(-1)^kn!}{(k!)^2(n-k)!}x^k.
\end{eqnarray}
One can check that the Theorem~\ref{thm density matrix} is fulfilled.
\section{Exponential family of pdfs}\label{sec_4}
As we can see from the example of harmonic oscillators, the pdfs \eqref{1004} and \eqref{1100} can be written in the following form:
\begin{eqnarray}
    W(X|\eta) \equiv h(X)e^{\eta^T \tau(X)-A(\eta)},
\end{eqnarray}
that  is a general expression for an exponential family of pdfs 
%whose density (relative to the vector of parameters $\eta$) have the following general form:
for a given vector of sufficient statistics $\tau(X)$ and normalization function $h(X)$.
For example, for the ground state of the harmonic oscillator, the functions are the following:
%\textcolor{blue}{also h doesn't seem to depend on X really or am I missing something here? I think we need to introduce $g(\mu, \nu)$ in someway before naming it here, maybe put $h(X,g(\mu, \nu)) = \frac{1}{\sqrt{\pi g(\mu, \nu)}} $ this doesn't seem like a general form though. maybe find something more like $h(\alpha, g(\mu, \nu) \propto \alpha g(\mu, \nu)$, also not sure about the minus sign in front of $g(\mu, \nu)$? } 
$\tau(X)= -X^2$, $\eta=-1/(\mu^2+\nu^2)$, $h(X)=1/\sqrt{\pi}$. 
\par 
From the normalization condition, we can find the cumulant generating function:
%\textcolor{blue}{via the Laplace transform of $h(X)$}:
\begin{eqnarray}\label{1254}
    e^{A(\eta)}=\int h(X) e^{\eta^T \tau(X)}dX.
    \label{eqn: Laplace transform}
    \end{eqnarray}
This shows that $A(\eta)$ is not a degree of freedom in the specification of an exponential family density. It is determined once $\eta$, $\tau(X)$ and $h(X)$ are determined:
    \begin{eqnarray}
  A(\eta)=\log{\left(\int h(X) e^{\eta^T \tau(X)}dX\right)}.
\end{eqnarray}
The characteristic function \eqref{eqn pdf to charac} in the point $t=1$ for such family of pdfs can be written as follows:
\begin{eqnarray}\label{1123}
    \phi(1;\eta(\mu,\nu))=\int e^{iX}   W(X|\eta)dX=e^{-A(\eta(\mu,\nu))}\int e^{iX+\eta(\mu,\nu)^T \tau(X)}  h(X)dX,
\end{eqnarray}
where $\eta(\mu,\nu)$ is a vector, depending on $\mu$, $\nu$.
If one wants to check the known pdf from the exponential family to generate the quantum state, one needs to check that the latter characteristic function satisfies the Theorem~\ref{thm density matrix}. We deduce the following:
\begin{cor}
\label{thm density matrix_2}
The characteristic function from an exponential family of pdfs \eqref{1123} describes a quantum state defined by a density matrix $\hat{\rho}$  if:
 \begin{eqnarray}\label{th1_1}
\int e^{iX+\eta(0,0)^T \tau(X)}  h(X)dX=e^{A(\eta(0,0))},
 \end{eqnarray}
  \begin{eqnarray}\label{th2_1}
0&\leq&\frac{1}{2\pi}  \iint \limits_{-\infty}^{\infty}e^{-A(\eta(\mu,\nu))-A(\eta(-\mu,-\nu))} \int e^{iX_1+\eta(\mu,\nu)^T \tau(X_1)}  h(X_1)dX_1\\\nonumber
 &\times&\int e^{iX_2+\eta(-\mu,-\nu)^T \tau(X_2)}  h(X_2)dX_2d\mu d\nu\leq 1,
 \end{eqnarray}
  \begin{eqnarray}\label{th3_2}
&&e^{A(\eta(\mu,y'-y))}\int e^{iX+\eta(\mu,y-y')^T \tau(X)}  h(X)dX\\\nonumber
&=& e^{A(\eta(\mu,y-y'))}\int e^{-iX+\eta(\mu,y'-y)^T \tau(X)}  h(X)dX,\quad \forall y,y',
\end{eqnarray}
\begin{eqnarray}
   \frac{1}{2\pi}\int e^{-A(\eta(\mu,0))}\int e^{iX+\eta(\mu,0)^T \tau(X)}  h(X)e^{-i\mu y}dX  d\mu\geq 0,\quad \forall y.
    \label{eqn: integral from matrix elements_2}
\end{eqnarray}
\end{cor}
\par Hence, any pdf from the exponential family can be checked on the latter conditions. Further, we provide more detailed analyses for the most important pdfs known in probability theory. 
\subsection{Special cases}
Let us observe $\tau(X)=X$, $X\in \mathbb{R}^+$, $A(\eta)$ is the $\log{}$ of  Laplace transform of $h(X)$. The Laplace transform of $h(X)$ is defined as follows:
\begin{eqnarray}
   H(s) =\int\limits_{0}^{\infty} h(X) e^{-sX}dX,
\end{eqnarray}
where $s$ is a complex frequency domain parameter
$ s=\sigma +i\omega$
with $\sigma, \omega \in \mathbb{R}$.
 A necessary condition for existence of the integral is that $h(X)$ must be locally integrable on $(0,\infty)$ and $\eta<0$. 
 Then the characteristic function in $t=1$ is
  \begin{eqnarray}
    \phi(1;\eta)=\left(\int\limits_{0}^{\infty} h(X_1) e^{\eta X_1}dX_1\right)^{-1}\left(\int\limits_{0}^{\infty} h(X_2) e^{(i+\eta ) X_2}  dX_2\right).
    \end{eqnarray}
\begin{exm}
Let us observe $h(X)=X^{\alpha-1}$, $\alpha>0$ and $\eta=-p(\mu,\nu)$, where $p(\mu,\nu)$ is a nonnegative function. Then the pdf is 
    \begin{eqnarray}
 f(X|\mu,\nu)=X^{\alpha-1}e^{-p(\mu,\nu) X-A(p(\mu,\nu))},\quad X\in \mathbb{R}^+.   
\end{eqnarray} 
Then the characteristic function is
  \begin{eqnarray}
    \phi(1;\mu,\nu)=
  %  \frac{\Gamma(\alpha)(p(\mu,\nu)-i)^{-\alpha}}{\Gamma(\alpha)p^{-\alpha}(\mu,\nu)}
   \frac{p^{\alpha}(\mu,\nu)}{(p(\mu,\nu)-i)^{\alpha}}.
    \end{eqnarray}
One can easily check that the latter type of functions does not satisfy the first condition of the Theorem~\ref{thm density matrix}. Hence, this class of pdfs is not suitable for a tomogram. This class includes important pdfs like exponential, gamma, and $\chi^2$.
\end{exm}

Let us observe $\tau(X)=(X,X^2)^T$, $\eta=(\eta_1,\eta_2)$, $X\in R$. Then the characteristic function is
     \begin{eqnarray}
    \phi(1;\eta)=\left(\int\limits_{-\infty}^{\infty}  h(X_1) e^{\eta_1 X_1+\eta_2 X_1^2}dX_1\right)^{-1}\left(\int\limits_{-\infty}^{\infty}  h(X_2) e^{(i+\eta_1)X_2+\eta_2X^2_2}  dX_2\right).
    \end{eqnarray}
    \begin{exm}
For a special case of $h(X)=C$, where $C$ is a constant and $\eta_1=p_1(\mu,\nu)$, $\eta_2=-p_2(\mu,\nu)$, where $p_i(\mu,\nu)$, $i=1,2$ are nonnegative functions, the characteristic function is the following:
    \begin{eqnarray}
    \phi(1;p(\mu,\nu))=%\left(\int\limits_{-\infty}^{\infty}   e^{\eta_1 X_1+\eta_2 X_1^2}dX_1\right)^{-1}\left(\int\limits_{-\infty}^{\infty}   e^{(i+\eta_1)X_2+\eta_2X^2_2}  dX_2\right)=
    e^{-\frac{2ip_1(\mu,\nu)-1}{4p_2(\mu,\nu)}}.
\end{eqnarray}
If we take $p_1(\mu,\nu)=0$, $p_2(\mu,\nu)=-1/(\mu^2+\nu^2)$ the case corresponds to harmonic oscillator \eqref{1407} and  the  tomogram is the Gaussian distribution.
%\par Let us check the Theorem~\ref{thm density matrix} conditions. 
%The first condition provides two possibilities:
%\begin{eqnarray}
%    \eta_1(0,0)=-\frac{i}{2},\quad \text{or}\quad \eta^{-1}_2(0,0)=0.
%\end{eqnarray}
%The second condition is
%  \begin{eqnarray}
% 0\leq\quad \frac{1}{2\pi} \iint \limits_{-\infty}^{\infty} e^{-\frac{2i\eta_1(\mu,\nu)-1}{4\eta_2(\mu,\nu)}}  e^{-\frac{2i\eta_1(-\mu,-\nu)-1}{4\eta_2(-\mu,-\nu)}}d\mu d\nu\leq 1.
% \end{eqnarray}
% The third is 
%\par Let us observe $h(X)=C X^n$, where $n$ is integer, and $\eta_1=p_1(\mu,\nu)$, $\eta_2=-p_2(\mu,\nu)$, where $p_i(\mu,\nu)$, $i=1,2$ are nonnegative function. Then the pdf is
% \begin{eqnarray}
%     f(X|\mu,\nu)=C X^n e^{-p_2(\mu,\nu)X^2+p_1(\mu,\nu)X-A(p(\mu,\nu))}.
% \end{eqnarray}
% We use the integral
%\begin{eqnarray}
%    \int\limits_{-\infty}^{\infty}x^n e^{-p x^2-q x}dx=\left(\frac{i}{2}\right)^n\sqrt{\pi}p^{-(n+1)/2}\exp{\left(\frac{q^2}{4p}\right)}H_n\left(\frac{iq}{2\sqrt{p}}\right)
%\end{eqnarray}
%to find the characteristic function
%    \begin{eqnarray}\label{1415}
%    \phi(1|\eta(\mu,\nu))=\frac{\exp{\left(\frac{2ip_1(\mu,\nu)-1}{4p_2(\mu,\nu)}\right)}H_n\left(\frac{ip_1(\mu,\nu)-1}{2\sqrt{p_2(\mu,\nu)}}\right)}{H_n\left(\frac{ip_1(\mu,\nu)}{2\sqrt{p_2(\mu,\nu)}}\right)}.
%    \end{eqnarray}
    \end{exm}

\section{Pseudoharmonic Oscillator}\label{sec_3}
\par 
%In this section, we draw our attention to  the Pseudoharmonic Oscillator (PHO)~\cite{olendski2023one}. 
The straight motion along the positive $x-$ semi-axis can be described by potential
%\liubov{change $\alpha$ to $a$ since coherent states matches}
\begin{eqnarray}\label{1548}
    V^{1D}(a,x)=D_{\omega}\left(\frac{x}{x_{\omega}}-\sqrt{a}\frac{x_{\omega}}{x}\right)^2,\quad 0\leq x\leq \infty,
\end{eqnarray}
where the dimensionless coefficient $a\geq 0$, $D_{\omega}=\frac{\hbar\omega}{2}$, $\omega$ is the confining frequency and $x_{\omega}=\frac{\hbar}{m\omega}$.
At $a= 0$, equation \eqref{1548} describes the geometry of a particle confined to the right-hand
half of the harmonic oscillator of the frequency $\omega$. The problem \eqref{1548} is cited in literature as pseudoharmonic oscillator (PHO)~\cite{olendski2023one}.
The 1D Schr{\"o}dinger equation is
\begin{eqnarray}
    \frac{\hbar^2}{2m}\frac{d^2}{dx^2}\Psi_n(a;x)+ V^{1D}(a,x)\Psi_n(a;x)= E_n(a)\Psi_n(a;x).
\end{eqnarray}
For the $a$-dependent potential \eqref{1548} the energies and the wave function are
\begin{eqnarray}
&&E_n(a)=\hbar\omega(2n+1+b-\sqrt{a}),\\\nonumber
&&    \Psi_n(a;x)=\frac{1}{\sqrt{x_{\omega}}}\left[\frac{2n!}{\Gamma(n+b+1)}\right]^{\frac{1}{2}} \left(\frac{x}{x_{\omega}}\right)^{b+\frac{1}{2}}\exp{\left(-\frac{1}{2}\frac{x^2}{x^2_{\omega}}\right)}L_n^{(\eta)}
    \left(\frac{x^2}{x^2_{\omega}}\right),
    \label{pseudoharmonic Oscillator energy and wave function}
\end{eqnarray}
where 
$b=\frac{1}{2}\sqrt{1+4a}$, $\Gamma(z)$ is a 
%\textcolor{blue}{I think when we refer to the Gamma function in introduction in summary we should at least put the word once in the section where it is discussed apart from graphic description} 
$\Gamma$-function, 
$L_n^{(b)}$ is an $n$th order associated Laguerre polynomial. The divergence of the potential  at the left edge $x = 0$
forces the function to vanish there at any $a$ and $n$: 	$ \Psi_n(a;0)=0$.
At $a = 0$ the spatial
dependence reads
\begin{eqnarray}\label{1627}
&&    \Psi_n(0;x)=\frac{(-1)^n}{x_{\omega}^{\frac{1}{2}}}\frac{1}{2^{2n+\frac{1}{2}}}\left[\frac{1}{n!\Gamma(n+\frac{3}{2})}\right]^{\frac{1}{2}} \exp{\left(-\frac{1}{2}\frac{x^2}{x^2_{\omega}}\right)}H_{2n+1}
    \left(\frac{x}{x_{\omega}}\right).
    \label{ground state HO}
\end{eqnarray}
Then, one can find the tomogram $ \mathcal{W}_n(X|a,\mu,\nu)$ of \eqref{1627}:
\begin{eqnarray}
  \mathcal{W}_n(X|0,\mu,\nu)\!\!&=&\!\!\frac{1}{2\pi|\nu|} \frac{1}{x_{\omega}}\frac{1}{2^{4n+1}}\frac{1}{n!\Gamma(n+\frac{3}{2})} I(X),\\\nonumber
  I(X)&\equiv& \Bigg|
  \int\limits_{0}^{\infty}
  \exp{\left[-x^2\left(\frac{1}{2x^2_{\omega}}-\frac{i\mu}{2\nu}\right)-\frac{iX}{\nu}x\right]}
  H_{2n+1}
    \left(\frac{x}{x_{\omega}}\right) dx\Bigg|^2.
    \label{tomogram integralform pseudoharmonic osc}
\end{eqnarray}
%\liubov{Is $X$ on $R$ or $R^+$?}
The closed form of the latter tomogram is derived in Appendix~\ref{ap_3}:
\begin{figure}[h]
  \center{\includegraphics[width=.8 \linewidth]{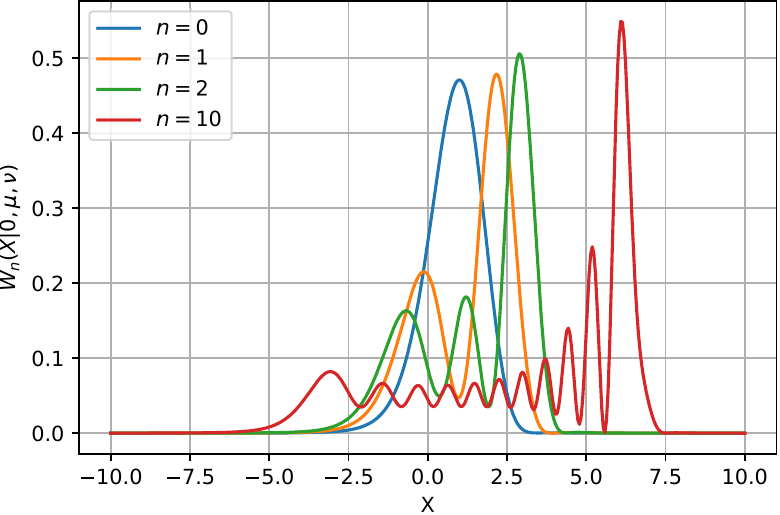}}
    \caption{The tomogram $\mathcal{W}_n(X|0,\mu,\nu)$ for the PHO  \eqref{1627} is plotted for $n\in\{0, 1, 2, 10\}$. The ground state tomogram is a Gaussian pdf with the shifted center, while the excited states are the multi-mode pdfs from the exponential family.}
    \label{fig:n=0,1,2,3}
\end{figure}
\begin{eqnarray}
   && \mathcal{W}_n(X|0,\mu,\nu)\!\!=\!\!
    \frac{1}{\pi} \frac{( (2n+1)!)^2 }{n!(n+\frac{1}{2})!} \frac{\nu^{2n+1} x_{\omega}}{(\nu^2+\mu^2 x^4_{\omega})^{n+1}} e^{-\frac{X^2 x^2_{\omega}}{2\left(\nu^2+\mu^2x^4_{\omega}\right)}}
\\\nonumber
 &\times& \Bigg[\sum_{m_1,m_2=0}^{n}
  \frac{(-1)^{m_1+m_2}}{{m_1}!{m_2}!} \left(\frac{x^2_\omega}{16\nu}\right)^{{m_1}+m_2} \left(\nu-i\mu x^2_{\omega}\right)^{m_1} 
 \left(\nu+i\mu x^2_{\omega}\right)^{m_2}  \\\nonumber&\times& D_{-(2n- 2{m_1}+2)}\left(\frac{iX x_{\omega}\sqrt{\nu+i\mu x^2_{\omega}}}{\sqrt{\nu}\sqrt{\nu^2+\mu^2 x^4_{\omega}}}\right)D_{-(2n- 2{m_2}+2)}\left(\frac{iX x_{\omega}\sqrt{\nu-i\mu x^2_{\omega}}}{\sqrt{\nu}\sqrt{\nu^2+\mu^2 x^4_{\omega}}}\right) 
  \Bigg].
  \label{eqn general central pseudo pdf_2}
\end{eqnarray}
This pdf is  a, so called, normal-exponential-gamma distribution type.
%For our purposes, it is insightful to look at plots of the pdf.  Using the representation we derived \ref{eqn general central pseudo pdf_2} we can get plots for arbitrary excited states of a centered ($a = 0$) pseudoharmonic oscillator. 
It is depicted in Fig.~\ref{fig:n=0,1,2,3} for the excitations $n = \{0,1 , 2, 10\}$. The distribution is not symmetrical with the center of the pdf shifted to the right side of the 
%\textcolor{blue}{I see the connection here but $x-$axis? definitely singular though right? axIs?} 
$X$-axis, whereas the pdf of a standard harmonic oscillator is centered at $X=0$. The characteristic function of the latter distribution defined by transformation \eqref{eqn pdf to charac_1} is quite cumbersome. We have numerically verified that it satisfies the conditions of Theorem~\ref{thm density matrix}. 
%When looking at the peaks from left to right we can see a rise in height, which is also recognizable for the standard HO going from the center peaks outwards but a lot less significant.
\par Next we move to the tomogram for the general case \eqref{pseudoharmonic Oscillator energy and wave function}. According to the definition it  can be written as
\begin{eqnarray}\label{1645_1}
    \mathcal{W}(X|a,\mu,\nu)=\frac{1}{2\pi|\nu|}\frac{1}{x_{\omega}}\left[\frac{2n!}{\Gamma(n+b+1)}\right]\Bigg|\int\limits_{0}^{\infty}  \left(\frac{y}{x_{\omega}}\right)^{b+\frac{1}{2}}e^{\left(-\frac{1}{2}\frac{y^2}{x^2_{\omega}}\right)}L_n^{(b)}
    \left(\frac{y^2}{x^2_{\omega}}\right)
    e^{(\frac{i\mu}{2\nu}y^2-\frac{iX}{\nu}y)} dy\Bigg|^2.
\end{eqnarray}

\begin{figure}[h]
\begin{minipage}[h]{0.49\linewidth}
\center{\includegraphics[width=0.9\linewidth]{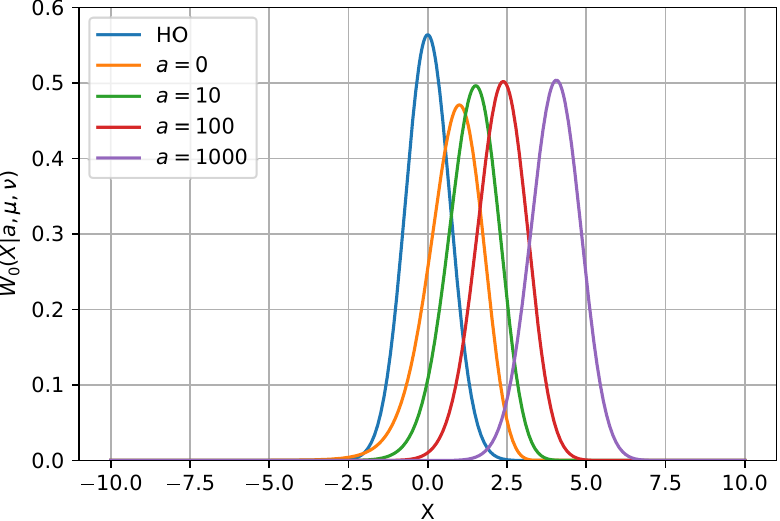} \\ a)}
\end{minipage}
\hfill
\begin{minipage}[h]{0.49\linewidth}
\center{\includegraphics[width=0.9\linewidth]{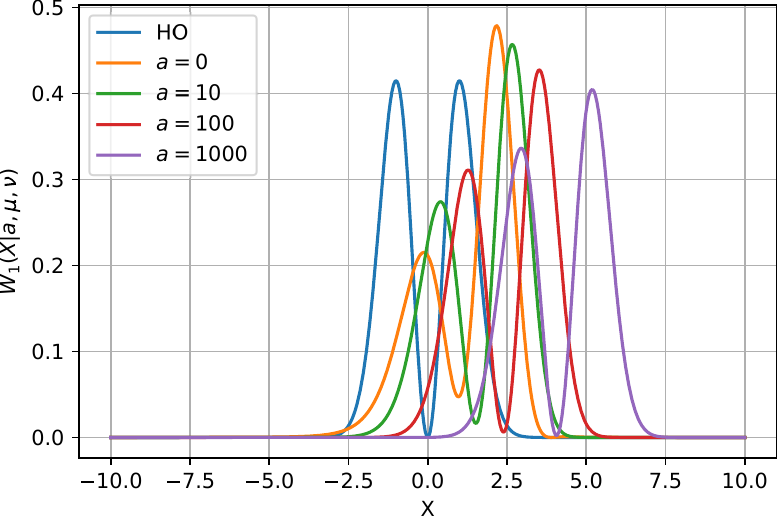} \\ b)}
\end{minipage}
\caption{Comparison of the tomograms of HO and the PHO for $a\in\{0,10,100,1000\}$, a) The ground state ($n=0$) tomograms are the Gaussian pdfs, where for the PHO the center of the pdf is shifted. b) the first excited state ($n=1$) tomograms are from the exponential family of pdf with multiple modes.}
\label{fig_2}
\end{figure}
The closed form is
\begin{eqnarray}\label{1327_2}
   &&  \mathcal{W}_n(X|a,\mu,\nu)=\frac{x_{\omega}}{2\pi|\nu|}\left[\frac{2n!}{\Gamma(n+b+1)}\right]\exp{\left(-\frac{X^2 x^2_{\omega}}{2(\nu^2+\mu^2 x^4_{\omega})}\right)}\left(\frac{(b+1)_n}{n}\right)^2\\\nonumber
&\times&     \Bigg| \sum\limits_{k=0}^{\infty}\frac{(n)_k}{(b+1)_k k!}\Gamma\left(2k+b+\frac{3}{2}\right)\left(\frac{\nu-i\mu x^2_{\omega}}{\nu}\right)^{-(k+\frac{b}{2}+\frac{3}{4})}   
 D_{-(2k+b+\frac{3}{2})}\left(\frac{iXx_{\omega}\sqrt{\nu+i\mu x^2_{\omega}} }{\sqrt{\nu}\sqrt{\nu^2+\mu^2x^4_{\omega}}}\right)\Bigg|^2.
 \label{eqn general pseudo shiftable pdf_1}
\end{eqnarray}
Using the latter representation, we can examine the dependence of $ \mathcal{W}_n(X|a,\mu,\nu)$ from $a$ and  compare it to the $ \mathcal{W}_n(X|\mu,\nu)$ pdf. Both functions are depicted in Fig.~\ref{fig_2} for $n=\{0,1\}$. We can observe that the larger the value of $a$ becomes, the further the function $ \mathcal{W}_n(X|a,\mu,\nu)$ is shifted to the right. The shape of the pdf changes slightly with the growth of $a$. However, for the first exited state, the mass of the picks of the distribution is slightly changing. 
\par We observe that,  both for HO and PHO, the tomograms deviate from Gaussian behavior even for the first excited state. The distribution becomes multi-modal and challenging to estimate. Subsequently, we will examine the case of an even more complex distribution arising in superposition states.

\section{Crystallized cat states}\label{sec_5}
The even and odd coherent states are introduced in \cite{dodonov1974even} from the Glauber coherent states $\ket{\alpha}$ of the harmonic oscillator as:
\begin{eqnarray}
   \ket{\alpha_{\pm}}=N_{\pm}(\ket{\alpha}\pm\ket{-\alpha}),\quad N_{\pm}=\left(2(1\pm e^{-2|\alpha|^2})\right)^{-1/2}.
\end{eqnarray}
Their generalization is provided in \cite{DODONOV1974597}, using the Abelian symmetry group $C_3$ with three rotation elements $\{1, e^{2\pi i/3},  e^{4\pi i/3}\}$, acting on the coherent states, giving the following state:
\begin{eqnarray}\label{1642}
    \ket{\psi}=N\sum\limits_{j=1}^{3} \ket{\psi_j},\quad  \ket{\psi_1}=\ket{\alpha},\quad  \ket{\psi_2}=\ket{\alpha e^{2\pi i/3}},\quad  \ket{\psi_3}=\ket{\alpha e^{4\pi i/3}}.
\end{eqnarray}
The coordinate representation of the coherent state reads as
\begin{eqnarray}
    \psi_{\alpha}(x)=\langle x|\alpha\rangle=\pi^{-1/4}\exp{\left(-\frac{x^2}{2}-\frac{|\alpha|^2}{2}+\sqrt{2}\alpha x-\frac{\alpha^2}{2}\right)},
\end{eqnarray}
and its tomogram is known to be equal to~\cite{man1997quantum}
\begin{eqnarray}\label{1300}
   \mathcal{W}(X|{\alpha},\mu,\nu)=  \frac{e^{-|\alpha|^2}e^{\frac{
(\nu+i\mu)^2\alpha^2 +(\nu-i\mu)^2(\alpha^{\star})^2}{2(\nu^2+\mu^2)}}}{\sqrt{\pi(\nu^2+\mu^2)}}e^{-\frac{X^2}{\nu^2+\mu^2}}e^{\frac{\sqrt{2}iX\left(
(\nu-i\mu)\alpha^{\star} -(\nu+i\mu)\alpha \right)}{\nu^2+\mu^2}}.
\end{eqnarray}
The characteristic function is
\begin{eqnarray}\label{1300_1}
   {\phi}(1;{\alpha},\mu,\nu)=  e^{-\frac{\nu^2+\mu^2}{4}}e^{-\frac{(\nu-i\mu)\alpha^{\star}-(\nu+i\mu)\alpha}{\sqrt{2}}}.
\end{eqnarray}
We can write every state in \eqref{1642} as
\begin{eqnarray}
   && \psi_{j}(x)= N_j\exp{\left(-A_j x^2+B_j x+C_j\right)},\quad j=1,2,3,\\\nonumber
    &&A_j=\frac{1}{2},\quad B_j=\sqrt{2}\alpha e^{\frac{2\pi i}{3}(j-1)},\quad C_j=-\frac{|\alpha|^2}{2}-\frac{\alpha^2e^{\frac{4\pi i}{3}(j-1)}}{2},\quad N_j=\pi^{-1/4}.
\end{eqnarray}
According to \eqref{1645_5} the tomogram can be written as
\begin{eqnarray}\label{1328}
    \mathcal{W}_{ccat}(X|\alpha,\mu,\nu)
&=&\frac{|N|^2e^{-|\alpha|^2}}{\sqrt{\pi(\nu^2+\mu^2)}}e^{-\frac{X^2}{\nu^2+\mu^2}}\\\nonumber
&\times&\sum\limits_{j,k=1}^{3}\exp{\left(\frac{\sqrt{2}iX\left(
(\nu-i\mu)\alpha^{\star} e^{-\frac{2\pi i}{3}(k-1)}-(\nu+i\mu)\alpha e^{\frac{2\pi i}{3}(j-1)}\right)}{\nu^2+\mu^2}\right)}\nonumber\\
&\times&\exp{\left(\frac{\left(
(\nu+i\mu)^2\alpha^2 e^{\frac{4\pi i}{3}(j-1)}+(\nu-i\mu)^2(\alpha^{\star})^2 e^{-\frac{4\pi i}{3}(k-1)}\right)}{2(\nu^2+\mu^2)}\right)}\nonumber.\end{eqnarray}
One can see that if $k,j=1$ the tomogram coincides with the single coherent state tomogram  \eqref{1300}. The pdf \eqref{1328} is a mixture of the Gaussian distributions~\cite{robertson1969some}. Note that the sum of Gaussian pdfs in general is not a Gaussian pdf itself. 
%\begin{eqnarray}
%&&e^{C_j+C_k^{\star}}=e^{-|\alpha|^2-\frac{1}{2}\left(\alpha^2e^{\frac{4\pi i}{3}(j-1)}+(\alpha^{\star})^2e^{-\frac{4\pi i}{3}(k-1)}\right)}
%\\   && B_j=\sqrt{2}\alpha e^{\frac{2\pi i}{3}(j-1)},
 %   B_k^{\star}=\sqrt{2}\alpha^{\star} e^{-\frac{2\pi i}{3}(k-1)}\\
 %   && B_j- B_k^{\star}=\sqrt{2}(\alpha e^{\frac{2\pi i}{3}(j-1)}
%    -\alpha^{\star} e^{-\frac{2\pi i}{3}(k-1)})\\
%    && B_j+ B_k^{\star}=\sqrt{2}(\alpha e^{\frac{2\pi i}{3}(j-1)}
%    +\alpha^{\star} e^{-\frac{2\pi i}{3}(k-1)})\\
%     && B_j^2+ B_k^{2\star}=2\alpha^2 e^{\frac{4\pi i}{3}(j-1)}+2\alpha^{2\star} e^{-\frac{4\pi i}{3}(k-1)}\\
%      && B_j^2-B_k^{2\star}=2\alpha^2 e^{\frac{4\pi i}{3}(j-1)}-2\alpha^{2\star} e^{-\frac{4\pi i}{3}(k-1)}\\
%\end{eqnarray}
\par The characteristic function is the following:
\begin{eqnarray}
   \phi_{ccat}(1; \alpha,\mu,\nu)&=&|N|^2e^{-\frac{\nu^2+\mu^2}{4}}\sum\limits_{j,k=1}^{3}e^{-|\alpha|^2(\exp{(-\frac{2\pi i}{3}(k-j)}+1)}\\\nonumber
&\times&  e^{-\frac{(\nu-i\mu)\alpha^{\star}\exp{(-\frac{2\pi i}{3}(k-1))}-(\nu+i\mu)\alpha \exp{(\frac{2\pi i}{3}(j-1))}}{\sqrt{2}}}.
\end{eqnarray}
One can check that it satisfies the Theorem~\ref{thm density matrix}. 
%\textcolor{blue}{again did we? if so we might put both in the appendix maybe? otherwise not...}. 
%I did it numerically, while it was your part of work to be honest :). 
\par The Gottesman-Kitaev-Preskill (GKP) code is specifically designed for continuous-variable quantum systems, and its states are characterized by coherent superpositions~\cite{grimsmo2021quantum}. In particular, the GKP code employs states that are approximate eigenstates of position and momentum operators. The GKP states are typically defined as Gaussian wavepacket superpositions, and they serve as an encoding for quantum information. Since the GKP code space is an abstract construction and the code words are not normalised, in practise the coherent state approximation is used to prepare the state lying entirely in the GKP code space~\cite{PhysRevLett.112.120504}.  
Thus, the tomogram of a specific superposition of coherent states serves as an approximation to the tomogram of the theoretical GKP state. We see in this a potential extension of the article into the realm of error correction codes.

%\section{GKP states, N00N states}

%\section{}
%\liubov{Entangled states theorem, Probability function}
\begin{table}[]
\begin{tabular}{|l|l|}
 \hline
\multicolumn{1}{|c|}{\begin{tabular}[c]{@{}c@{}}Quantum \\ state\end{tabular}} & \multicolumn{1}{c|}{\begin{tabular}[c]{@{}c@{}}Tomogram (pdf)\\ Characteristic function\end{tabular}}  \\ \hline
HO (ground state)                                                                 &                                                                    $
     \mathcal{W}_0(X|\mu,\nu) =
     \frac{1}{\sqrt{\pi(\mu^2+\nu^2)}}\exp{\left[-\frac{X^2}{\mu^2+\nu^2}\right]}$,
                                                                                \\ 
                                                                 &                                                                    
$ {\phi}_0(1|\mu,\nu)=\exp{\left[-\frac{\mu^2+\nu^2}{4}\right]}$                                                                                    \\   
    \hline 
                                                                            HO (excited state)     &   $\mathcal{W}_n(X|\mu,\nu) =
       \mathcal{W}_0(X|\mu,\nu)\frac{1}{2^n n!}\Big(H_n\left(\frac{X}{\sqrt{\mu^2+\nu^2}}\right)\Big)^2$                                                                                                                                            \\ 
    &    $ {\phi}_n(1;\mu,\nu)
    ={\phi}_0(1|\mu,\nu) L_n\left(\frac{\mu^2+\nu^2}{2}\right)$                                                                                                                                                                \\ \hline 
                   PHO ($a=0$)                                                           &                                                   $
    \mathcal{W}_n(X|\frac{1}{2},\mu,\nu)\!\!=\!\!
    \frac{1}{\pi} \frac{( (2n+1)!)^2 }{n!(n+\frac{1}{2})!} \frac{\nu^{2n+1} x_{\omega}}{(\nu^2+\mu^2 x^4_{\omega})^{n+1}} e^{-\frac{X^2 x^2_{\omega}}{2\left(\nu^2+\mu^2x^4_{\omega}\right)}}
$\\
&
 $\times\Bigg[\sum_{m_1,m_2=0}^{n}
  \frac{(-1)^{m_1+m_2}}{{m_1}!{m_2}!} \left(\frac{x^2_\omega}{16\nu}\right)^{{m_1}+m_2} \left(\nu-i\mu x^2_{\omega}\right)^{m_1} 
 \left(\nu+i\mu x^2_{\omega}\right)^{m_2} $\\
 &$ \times D_{-(2n- 2{m_1}+2)}\left(\frac{iX x_{\omega}\sqrt{\nu+i\mu x^2_{\omega}}}{\sqrt{\nu}\sqrt{\nu^2+\mu^2 x^4_{\omega}}}\right)D_{-(2n- 2{m_2}+2)}\left(\frac{iX x_{\omega}\sqrt{\nu-i\mu x^2_{\omega}}}{\sqrt{\nu}\sqrt{\nu^2+\mu^2 x^4_{\omega}}}\right) 
  \Bigg].
 $                                                                                                                \\ \hline 
PHO  ($\forall a$)                                                          &  $ \mathcal{W}_n(X|\eta,\mu,\nu)=\frac{x_{\omega}}{2\pi|\nu|}\left[\frac{2n!}{\Gamma(n+\eta+1)}\right]\exp{\left(-\frac{X^2 x^2_{\omega}}{2(\nu^2+\mu^2 x^4_{\omega})}\right)}\left(\frac{(\eta+1)_n}{n}\right)^2$\\
&$  \Bigg| \sum\limits_{k=0}^{\infty}\frac{(n)_k}{(\eta+1)_k k!}\Gamma\left(2k+\eta+\frac{3}{2}\right)\left(\frac{\nu-i\mu x^2_{\omega}}{\nu}\right)^{-(k+\frac{\eta}{2}+\frac{3}{4})}   
\!\!\! D_{-(2k+\eta+\frac{3}{2})}\left(\frac{iXx_{\omega}\sqrt{\nu+i\mu x^2_{\omega}} }{\sqrt{\nu}\sqrt{\nu^2+\mu^2x^4_{\omega}}}\right)\Bigg|^2$\\\hline
                   Coherent state                                                           &    $   \mathcal{W}(X|{\alpha},\mu,\nu)=    \mathcal{W}_0(X|\mu,\nu)e^{-|\alpha|^2}e^{\frac{
(\nu+i\mu)^2\alpha^2 +(\nu-i\mu)^2(\alpha^{\star})^2}{2(\nu^2+\mu^2)}}e^{\frac{\sqrt{2}iX\left(
(\nu-i\mu)\alpha^{\star} -(\nu+i\mu)\alpha \right)}{\nu^2+\mu^2}}$                                                                                                                                                                       \\ 
&${\phi}(1;{\alpha},\mu,\nu)=  {\phi}_0(1;\mu,\nu)e^{-\frac{(\nu-i\mu)\alpha^{\star}-(\nu+i\mu)\alpha}{\sqrt{2}}}$\\\hline 
Crystallized & $\mathcal{W}_{ccat}(X|\alpha,\mu,\nu)= \mathcal{W}_0(X|\mu,\nu)e^{-|\alpha|^2}|N|^2$\\
cat states&$\!\!\times\!\!\sum\limits_{j,k=1}^{3}e^{\frac{
(\nu+i\mu)^2\alpha^2 e^{\frac{4\pi i}{3}(j-1)}+(\nu-i\mu)^2(\alpha^{\star})^2 e^{-\frac{4\pi i}{3}(k-1)}}{2(\nu^2+\mu^2)}}\!\!e^{\frac{\sqrt{2}iX(
(\nu-i\mu)\alpha^{\star} e^{-\frac{2\pi i}{3}(k-1)}-(\nu+i\mu)\alpha e^{\frac{2\pi i}{3}(j-1)})}{\nu^2+\mu^2}}$\\
&${\phi}_{ccat}(1;{\alpha},\mu,\nu)= |N|^2{\phi}_0(1;\mu,\nu)$\\
&$\sum\limits_{j,k=1}^{3}e^{-|\alpha|^2(\exp{(-\frac{2\pi i}{3}(k-j)}+1)}
  e^{-\frac{(\nu-i\mu)\alpha^{\star}\exp{(-\frac{2\pi i}{3}(k-1))}-(\nu+i\mu)\alpha \exp{(\frac{2\pi i}{3}(j-1))}}{\sqrt{2}}}$\\
\hline 
\end{tabular}
\caption{Tomograms and characteristic functions for the ground and excited states of the oscillator and pseudoharmonic oscillator, as well as for coherent states and superpositions of cat states.}\label{tab_1}
\end{table}

\section{Summary}
\par For the accurate reconstruction of a quantum state, various techniques are employed. The tomographic pdf, being the true distribution function, completely characterizes the state and have potential to be a convenient method for quantum state reconstruction. When applying classical probability theory methods for estimating the pdf, it is often necessary to know which distribution family we are working with. In this article, we thoroughly investigated the conditions (Theorem~\ref{thm density matrix}) that the characteristic function of the distribution must satisfy to describe a physical system. This characteristic function is obtained as the Fourier transform of the pdf tomogram, aiming to describe a physical quantum state. These conditions strictly limit the family of pdfs suitable for describing quantum states.
%These conditions strictly limit the family of densities suitable for generating quantum states.
We specifically explored the exponential family of densities, introducing conditions that the characteristic function of a quantum state from this family must satisfy. We tested several well-known distributions, such as exponential, gamma, and $\chi^2$, demonstrating their unsuitability for generating quantum states.
\par Using the characteristic function, it is possible to express key features of a quantum state, e.g., the purity parameter. Additionally, we introduced a trace distance between two states based on their characteristic functions. Consequently, knowledge of the estimate of the characteristic function can serve as a test for the purity or proximity of two quantum states.
\par  We investigated well-known states such as the quantum oscillator (symmetric and asymmetric), coherent states, cat states, and their superpositions. The tomograms of these states are associated with Gaussian and Hermite-Gaussian distributions or mixtures of these distributions, as shown in Table~\ref{tab_1}. Their characteristic functions satisfy the Theorem~\ref{thm density matrix} as expected.
\par The study clearly demonstrates that the family of densities capable of generating quantum states is extremely limited. Further development of the article could involve a detailed examination of well-known density distributions, exploring their potential as tomograms for quantum states. Our intuition suggests that the presence of a quadratic exponential term (as in the tomography of the oscillator and all other tomograms in Table~\ref{tab_1}) is likely mandatory. It would be particularly interesting to investigate the power-law family of distributions, which finds extensive application in classical mathematics too. This study provides insights into the family of distributions encountered in quantum mechanics. With this information, one can employ both parametric and non-parametric methods to estimate the tomograms of quantum states. This investigation will be the focus of our future work.

\section{Acknowledgments}
L.M. was partly supported by the Netherlands Organisation for Scientific Research (NWO/OCW), as part of the Quantum Software Consortium program (project number 024.003.037 / 3368).  This research work was partly supported by the Roadmap for the Development of Quantum Technologies in Russian Federation, contract No. 868-1.3-15/15-2021.

\bibliographystyle{unsrt}
\bibliography{bib}

\appendix

\section{Matrix elements of the density matrix}\label{app_1}
To write the condition $\rho{(y,y')}=\rho(y',y)^{\star}$ in terms of the characteristic function, we use 
%the expression $ \exp{(i(X\hat{1}-\mu\hat{q}-\nu\hat{p}))}= \exp{(iX\hat{1})}\exp{(-i(\mu\hat{q}+\nu\hat{p}))}$. Using
the Baker- Campbell- Hausdorff formula $e^{ia\hat{q}} e^{ib\hat{p}} e^{\frac{iab\hbar}{2}}= e^{i \left(a\hat{q} + b\hat{p}\right)}$. We get:
\begin{eqnarray}
    \exp{(i(X\hat{1}-\mu\hat{q}-\nu\hat{p}))}= \exp{(iX)}\exp{(-i\mu\hat{q})}\exp{(-i\nu\hat{p})}\exp{(i\frac{\mu\nu}{2})}.
\end{eqnarray}
The matrix element of the latter is
\begin{eqnarray}
    \Big[\exp{(i(X\hat{1}-\mu\hat{q}-\nu\hat{p}))}\Big]_{yy'}&=& \exp{(iX)}\exp{(i\frac{\mu\nu}{2})}\Big[\exp{(-i\mu\hat{q})}\exp{(-i\nu\hat{p})}\Big]_{yy'}\\\nonumber
    &=&\exp{(iX)}\exp{(i\frac{\mu\nu}{2})}\Big[\exp{(-i\mu\hat{q})}\Big]_{yx}\Big[\exp{(-i\nu\hat{p})}\Big]_{xy'}.
\end{eqnarray}
The matrix element from the latter expression is
\begin{eqnarray}
  \Big[\exp{(-i\mu\hat{q})}\Big]_{yx}=\langle y|\exp{(-i\mu\hat{q})}|x\rangle=\exp{(-i\mu x)}  \delta(y-x).
\end{eqnarray}
The  translation operator 
$\hat  {T}(\nu)=\exp{(-i\nu \hat{p}/\hbar)}$ moves particles and fields by the amount $\nu$, namely 
\begin{eqnarray}&&e^{-i\nu \hat{p}}\Psi(x)=\Psi(x+\nu),\quad e^{-i\nu \hat{p}}|x\rangle=|x+\nu\rangle.
\end{eqnarray}
Thus, we have
\begin{eqnarray}
  \Big[\exp{(-i\nu\hat{p})}\Big]_{xy'}=\langle x|\exp{(-i\nu\hat{p})}|y'\rangle=\langle x|y'+\nu\rangle=  \delta(x-y'-\nu).
\end{eqnarray}
Using  \eqref{1311}, we can write the matrix element of the density matrix as follows
\begin{eqnarray}\label{1623}
    \rho(y,y')=\frac{1}{2\pi}\int \phi(1;\mu,y-y')\exp{\left(-i\frac{\mu(y+y')}{2}\right)}  d\mu.
    \label{eqn: integral from matrix elements_2}
\end{eqnarray}
The  conjugate transpose element is
\begin{eqnarray}
    \rho^{\star}(y',y)=\frac{1}{2\pi}\int \phi(-1;\mu,y'-y)\exp{\left(-i\frac{\mu(y+y')}{2}\right)}  d\mu.
\end{eqnarray}

\section{Tomogram for the pseudoharmonic oscillator with $a=0$.}\label{ap_3}
\par Let us first find the tomogram $ \mathcal{W}_n(X|a,\mu,\nu)$ of \eqref{1627}:
\begin{eqnarray}
  \mathcal{W}_n(X|0,\mu,\nu)\!\!&=&\!\!\frac{1}{2\pi|\nu|} \frac{1}{x_{\omega}}\frac{1}{2^{4n+1}}\frac{1}{n!\Gamma(n+\frac{3}{2})} I(X),\\\nonumber
  I(X)&\equiv& \Bigg|
  \int\limits_{0}^{\infty}
  \exp{\left[-x^2\left(\frac{1}{2x^2_{\omega}}-\frac{i\mu}{2\nu}\right)-\frac{iX}{\nu}x\right]}
  H_{2n+1}
    \left(\frac{x}{x_{\omega}}\right) dx\Bigg|^2.
    \label{tomogram integralform pseudoharmonic osc_2}
\end{eqnarray}
Using the series expression of the Hermitian polynomial
\begin{eqnarray}
    H_{2n+1}\left(\frac{x}{x_{\omega}}\right)  %= (2n+1)! \sum_{m=0}^{\left\lfloor \tfrac{2n+1}{2} \right\rfloor} \frac{(-1)^m}{m!(2n- 2m+1 )!} \left(\frac{2x}{x_{\omega}}\right) ^{2n- 2m+1 }
    = (2n+1)! \sum_{m=0}^{n} \frac{(-1)^m}{m!(2n- 2m+1 )!} \left(\frac{2x}{x_{\omega}}\right) ^{2n- 2m+1 },
\end{eqnarray}
%\liubov{we can write till $n$, no floor function. since ${\left\lfloor \tfrac{2n+1}{2} \right\rfloor}=n$.}
we can rewrite the integral in the tomogram as
\begin{eqnarray}
I(X)=  %( (2n+1)!)^2 \Big| \sum_{m=0}^{\left\lfloor \tfrac{2n+1}{2} \right\rfloor}
( (2n+1)!)^2 \Big| \sum_{m=0}^{n}
 \frac{(-1)^m}{m!(2n- 2m+1 )!} \int\limits_{0}^{\infty}
 e^{\left[-x^2\left(\frac{1}{2x^2_{\omega}}-\frac{i\mu}{2\nu}\right)-\frac{iX}{\nu}x\right]}
 \left(\frac{2x}{x_{\omega}}\right) ^{2n- 2m+1 }dx\Big|^2 . 
\label{int sub}
\end{eqnarray}
We use the integral
\begin{eqnarray}\label{1327}
    &&\int\limits_{0}^{\infty} x^{\alpha-1} e^{-px^2-qx}dx=\Gamma(\alpha)(2p)^{-\alpha/2}
    \exp{(q^2/8p)}D_{-\alpha}(q/\sqrt{2p}),\\\nonumber
    &&[\Re{(\alpha)},\Re{(p)}>0],\quad \text{or}\quad [\Re{(\alpha)},\Re{(q)}>0,\Re{(p)}=0],\\\nonumber
    &&\text{or}\quad [0<\Re{(\alpha)}<2,\Re{(q)}=\Re{(p)}=0,\Im{(p)}\neq 0],
\end{eqnarray}
where $D_a(z)$ is parabolic cylinder functions, to write\\
%\begin{eqnarray}
%I&=&( (2n+1)!)^2 \Bigg| e^{\frac{-X^2}{4\nu^2\left(\frac{1}{x^2_{\omega}}+\frac{i\mu}{\nu}\right)}}D_{-2n+ 2m-2}\left(\frac{iX}{\nu\sqrt{(\left(\frac{1}{x^2_{\omega}}+\frac{i\mu}{\nu}\right))}}\right) \\\nonumber
% &\times& \sum_{m=0}^{\left\lfloor \tfrac{2n+1}{2} \right\rfloor}
%  \frac{(-1)^m}{m!(2n- 2m+1 )!} \left(\frac{2}{x_{\omega}}\right) ^{2n- 2m+1 }\Gamma(2n-2m+2)\left(\frac{1}{x^2_{\omega}}+\frac{i\mu}{\nu}\right)^{-n+m-1}
%  \Bigg|^2 
%\end{eqnarray}
%\textcolor{blue}{alternatively we could use the following integrals
%\begin{eqnarray}
%    \int_{-\infty}^{\infty} x^n \exp{(-px^2 - qx)} \\
%    = (\frac{i}{2})^n \sqrt{\pi} p^{-\frac{n+1}{2}} \exp{(\frac{q^2}{4p})} H_n(\frac{iq}{2 \sqrt{p}})
%    \label{alternative integral 1}
%\end{eqnarray}}
\begin{eqnarray}
I(X)&=&%( (2n+1)!)^2  e^{-\frac{X^2 x^2_{\omega}}{2\left(\nu^2+\mu^2x^4_{\omega}\right)}}\Bigg|\sum_{m=0}^{\left\lfloor \tfrac{2n+1}{2} \right\rfloor}
( (2n+1)!)^2  e^{-\frac{X^2 x^2_{\omega}}{2\left(\nu^2+\mu^2x^4_{\omega}\right)}}\Bigg|\sum_{m=0}^{n}
  \frac{(-1)^m}{m!(2n- 2m+1 )!} \Gamma(2n-2m+2)\\\nonumber
 &\times& D_{-2n+ 2m-2}\left(\frac{iX x_{\omega}}{\sqrt{\nu}\sqrt{\nu-i\mu x^2_{\omega}}}\right) \left(\frac{2}{x_{\omega}}\right) ^{2n- 2m+1 }\left(\frac{\nu-i\mu x^2_{\omega}}{\nu x^2_{\omega}}\right)^{-n+m-1}
  \Bigg|^2 .
\end{eqnarray}
Then the  tomogram is
\begin{eqnarray}
   && \mathcal{W}_n(X|0,\mu,\nu)\!\!=\!\!
    \frac{1}{\pi} \frac{( (2n+1)!)^2 }{n!\Gamma(n+\frac{3}{2})} \frac{\nu^{2n+1} x_{\omega}}{(\nu^2+\mu^2 x^4_{\omega})^{n+1}} e^{-\frac{X^2 x^2_{\omega}}{2\left(\nu^2+\mu^2x^4_{\omega}\right)}}
    \\\nonumber
 &\times& \Bigg|%\sum_{m=0}^{\left\lfloor \tfrac{2n+1}{2} \right\rfloor}
 \sum_{m=0}^{n}
  \frac{(-1)^m}{m!(2n- 2m+1 )!} \Gamma(2n-2m+2)\\\nonumber
 &\times&D_{-(2n- 2m+2)}\left(\frac{iX x_{\omega}\sqrt{\nu+i\mu x^2_{\omega}}}{\sqrt{\nu}\sqrt{\nu^2+\mu^2 x^4_{\omega}}}\right) \left(\frac{(\nu-i\mu x^2_{\omega})}{4\nu }\right)^m \left(\frac{x_\omega}{2}\right)^{2m}
  \Bigg|^2 .
\end{eqnarray}
%\liubov{Since $-(2n-2m+1)$ can be negative, we cant use the Hermite polynomial representation}
We use the series representation 
\begin{eqnarray}
     D_{\nu}(z)=\frac{2^{-\frac{\nu}{2}}e^{-\frac{z^2}{4}}}{\Gamma(\frac{1}{2}+\nu)}{}_1F_1(-\nu,\frac{1}{2},\frac{1}{2}z^2),
\end{eqnarray}
where ${}_1F_1(a,b,z)$ is the confluent hypergeometric function of the first kind.
It is also known that
\begin{eqnarray}
{}_1F_1(a;b;z)=\sum\limits_{k=0}^{\infty}\frac{(a)_kz^k}{(b)_k k!}, 
\end{eqnarray}
where  $a_k$ is the Pochhammer symbol:
$(a)_k=a(a+1)\dots (a+k-1)$, $k=1,2,3\dots$, $(a)_0=1$. Combining this knowledge, we can write
\begin{eqnarray}
     D_{\nu}(z)=\frac{2^{-\frac{\nu}{2}}e^{-\frac{z^2}{4}}}{\Gamma(\frac{1}{2}+\nu)}\sum\limits_{k=0}^{\infty}\frac{(-\nu)_kz^{2k}}{{2^k}(\frac{1}{2})_k k!}=
     \frac{2^{-\frac{\nu}{2}}e^{-\frac{z^2}{4}}}{\Gamma(\frac{1}{2}+\nu)}\sum\limits_{k=0}^{\infty}\frac{\Gamma(-\nu+k)z^{2k}}{\Gamma(-\nu)(2k-1)!! k!},
\end{eqnarray}
where we used $(x)_n=\frac{\Gamma(x+n)}{\Gamma(x)}$ and $(\frac{1}{2})_n=\frac{(2n-1)!!}{2^n}$.
Then we can write
\begin{eqnarray}
     D_{-(2n- 2m+2)}\left(\frac{iX x_{\omega}\sqrt{\nu+i\mu x^2_{\omega}}}{\sqrt{\nu}\sqrt{\nu^2+\mu^2 x^4_{\omega}}}\right)&=&\frac{2^{n- m+1}e^{\frac{X^2 x^2_{\omega}(\nu+i\mu x^2_{\omega})}{4\nu(\nu^2+\mu^2 x^4_{\omega})})}}{\Gamma(\frac{1}{2}-(2n- 2m+2))}\\\nonumber
 &\times&\sum\limits_{k=0}^{\infty}\frac{\Gamma(2n- 2m+2+k)\left(\frac{-X^2 x^2_{\omega}(\nu+i\mu x^2_{\omega})}{\nu(\nu^2+\mu^2 x^4_{\omega})}\right)^{k}}{\Gamma(2n- 2m+2)(2k-1)!!k!},
\end{eqnarray}
holds. It is easy to see that the complex conjugate of the latter function is
\begin{eqnarray}
     D^{\star}_{-(2n- 2m+2)}\left(\frac{iX x_{\omega}\sqrt{\nu+i\mu x^2_{\omega}}}{\sqrt{\nu}\sqrt{\nu^2+\mu^2 x^4_{\omega}}}\right)
     %=\frac{2^{n- m+1}e^{\frac{X^2 x^2_{\omega}(\nu-i\mu x^2_{\omega})}{4\nu(\nu^2+\mu^2 x^4_{\omega})})}}{\Gamma(\frac{1}{2}-(2n- 2m+2))}\sum\limits_{k=0}^{\infty}\frac{\Gamma(2n- 2m+2+k)\left(\frac{-X^2 x^2_{\omega}(\nu-i\mu x^2_{\omega})}{\nu(\nu^2+\mu^2 x^4_{\omega})}\right)^{k}}{\Gamma(2n- 2m+2)(2k-1)!!k!}\\\nonumber
    &=& D_{-(2n- 2m+2)}\left(\frac{iX x_{\omega}\sqrt{\nu-i\mu x^2_{\omega}}}{\sqrt{\nu}\sqrt{\nu^2+\mu^2 x^4_{\omega}}}\right).
\end{eqnarray}
Then the tomogram is the following

\begin{eqnarray}
   && \mathcal{W}_n(X|0,\mu,\nu)\!\!=\!\!
    \frac{1}{\pi} \frac{( (2n+1)!)^2 }{n!(n+\frac{1}{2})!} \frac{\nu^{2n+1} x_{\omega}}{(\nu^2+\mu^2 x^4_{\omega})^{n+1}} e^{-\frac{X^2 x^2_{\omega}}{2\left(\nu^2+\mu^2x^4_{\omega}\right)}}
\\\nonumber
 &\times& \Bigg[\sum_{m_1,m_2=0}^{n}
  \frac{(-1)^{m_1+m_2}}{{m_1}!{m_2}!} \left(\frac{x^2_\omega}{16\nu}\right)^{{m_1}+m_2} \left(\nu-i\mu x^2_{\omega}\right)^{m_1} 
 \left(\nu+i\mu x^2_{\omega}\right)^{m_2}  \\\nonumber&\times& D_{-(2n- 2{m_1}+2)}\left(\frac{iX x_{\omega}\sqrt{\nu+i\mu x^2_{\omega}}}{\sqrt{\nu}\sqrt{\nu^2+\mu^2 x^4_{\omega}}}\right)D_{-(2n- 2{m_2}+2)}\left(\frac{iX x_{\omega}\sqrt{\nu-i\mu x^2_{\omega}}}{\sqrt{\nu}\sqrt{\nu^2+\mu^2 x^4_{\omega}}}\right) 
  \Bigg].
  \label{eqn general central pseudo pdf}
\end{eqnarray}

\section{Tomogram for the pseudoharmonic oscillator for any $a$ }\label{ap_4}
Let us find the  tomogram for the \eqref{pseudoharmonic Oscillator energy and wave function}. According to \eqref{1645_5} it can be written
\begin{eqnarray}\label{1645_3}
    \mathcal{W}(X|\mu,\nu)&=&\frac{1}{2\pi|\nu|}\frac{1}{x_{\omega}}\left[\frac{2n!}{\Gamma(n+\eta+1)}\right]\\\nonumber
    &\times&\Big|\int\limits_{0}^{\infty}  \left(\frac{y}{x_{\omega}}\right)^{\eta+\frac{1}{2}}\exp{\left(-\frac{1}{2}\frac{y^2}{x^2_{\omega}}\right)}L_n^{(\eta)}
    \left(\frac{y^2}{x^2_{\omega}}\right)
    \exp{(\frac{i\mu}{2\nu}y^2-\frac{iX}{\nu}y)} dy\Big|^2.
\end{eqnarray} 
%\liubov{one plot: $\alpha=0,1,10,100$, one plot checking that (58) and (46) are close. One plot normal oscillator $n=0,1$ and this oscillator $n=0,1$ }
%\textcolor{blue}{I believe we don't need the first plot but rather the second too fulfill our purpose}
Changing of variables to $p=\frac{1}{2}-\frac{i\mu}{2\nu}x^2_{\omega}$, $q=\frac{iX}{\nu} x_{\omega}$, we get
\begin{eqnarray}\label{1645}
    \mathcal{W}_n(X|\mu,\nu)=\frac{x_{\omega}}{2\pi|\nu|}\left[\frac{2n!}{\Gamma(n+\eta+1)}\right]\Big|\int\limits_{0}^{\infty}  t^{\eta+\frac{1}{2}}\exp{\left(-p t^2-qt\right)} L_n^{(\eta)}(
    t^2)dt\Big|^2.
\end{eqnarray}
Similarly to the previous subsection, we can use the confluent hypergeometric function representation:
\begin{eqnarray}
   L^{(\eta)
   }_n(t^2)=\frac{(\eta+1)_n}{n}{}_1F_1(-n;\eta+1;t^2) =\frac{(\eta+1)_n}{n} \sum\limits_{k=0}^{\infty}\frac{(n)_kt^{2k}}{(\eta+1)_k k!} .
\end{eqnarray}
Then the tomogram can be written as 
\begin{eqnarray*}
    \mathcal{W}_n(X|\mu,\nu)=\frac{x_{\omega}}{2\pi|\nu|}\left[\frac{2n!}{\Gamma(n+\eta+1)}\right]\Big|\frac{(\eta+1)_n}{n} \sum\limits_{k=0}^{\infty}\frac{(n)_k}{(\eta+1)_k k!}\int\limits_{0}^{\infty}  t^{2k+\eta+\frac{1}{2}}\exp{\left(-p t^2-qt\right)}  dt\Big|^2.
\end{eqnarray*}
We use \eqref{1327} to write the closed form of the tomogram
\begin{eqnarray}\label{1327_1}
   &&  \mathcal{W}_n(X|\mu,\nu)=\frac{x_{\omega}}{2\pi|\nu|}\left[\frac{2n!}{\Gamma(n+\eta+1)}\right]\exp{\left(-\frac{X^2 x^2_{\omega}}{2(\nu^2+\mu^2 x^4_{\omega})}\right)}\left(\frac{(\eta+1)_n}{n}\right)^2\\\nonumber
&\times&     \Bigg| \sum\limits_{k=0}^{\infty}\frac{(n)_k}{(\eta+1)_k k!}\Gamma\left(2k+\eta+\frac{3}{2}\right)\left(\frac{\nu-i\mu x^2_{\omega}}{\nu}\right)^{-(k+\frac{\eta}{2}+\frac{3}{4})}   
 D_{-(2k+\eta+\frac{3}{2})}\left(\frac{iXx_{\omega}\sqrt{\nu+i\mu x^2_{\omega}} }{\sqrt{\nu}\sqrt{\nu^2+\mu^2x^4_{\omega}}}\right)\Bigg|^2.
 \label{eqn general pseudo shiftable pdf}
\end{eqnarray}

\end{document}